\documentclass[aps,prl,twocolumn,superscriptaddress,showpacs]{revtex4}

\usepackage{graphicx}

\begin{document}

\title{Bandwidth-disorder phase diagram of half doped layered manganites}

\author{R. Mathieu\cite{newaddress}}
\affiliation{Spin Superstructure Project (ERATO-SSS), JST, AIST Central 4, Tsukuba 305-8562, Japan}

\author{M. Uchida}
\affiliation{Spin Superstructure Project (ERATO-SSS), JST, AIST Central 4, Tsukuba 305-8562, Japan}

\author{Y. Kaneko}
\affiliation{Spin Superstructure Project (ERATO-SSS), JST, AIST Central 4, Tsukuba 305-8562, Japan}

\author{J. P. He}
\affiliation{Spin Superstructure Project (ERATO-SSS), JST, AIST Central 4, Tsukuba 305-8562, Japan}

\author{X. Z. Yu}
\affiliation{Spin Superstructure Project (ERATO-SSS), JST, AIST Central 4, Tsukuba 305-8562, Japan}

\author{R. Kumai}
\affiliation{Correlated Electron Research Center (CERC), AIST Central 4, Tsukuba 305-8562, Japan}

\author{T. Arima}
\affiliation{Spin Superstructure Project (ERATO-SSS), JST, AIST Central 4, Tsukuba 305-8562, Japan}
\affiliation{Institute of Multidisciplinary Research for Advanced Materials, Tohoku University, Sendai 980-8577, Japan}

\author{Y. Tomioka}
\affiliation{Correlated Electron Research Center (CERC), AIST Central 4, Tsukuba 305-8562, Japan}

\author{A. Asamitsu}
\affiliation{Spin Superstructure Project (ERATO-SSS), JST, AIST Central 4, Tsukuba 305-8562, Japan}
\affiliation{Cryogenic Research Center (CRC), University of Tokyo, Bunkyo-ku, Tokyo 113-0032, Japan}

\author{Y. Matsui}
\affiliation{Spin Superstructure Project (ERATO-SSS), JST, AIST Central 4, Tsukuba 305-8562, Japan}
\affiliation{Advanced Materials Laboratory, National Institute for Materials Science (NIMS), Tsukuba 305-0044, Japan}

\author{Y. Tokura}
\affiliation{Spin Superstructure Project (ERATO-SSS), JST, AIST  Central 4, Tsukuba 305-8562, Japan}
\affiliation{Correlated Electron Research Center (CERC), AIST Central 4, Tsukuba 305-8562, Japan}
\affiliation{Department of Applied Physics, University of Tokyo, Tokyo 113-8656, Japan}

\date{\today}

\begin{abstract}
Phase diagrams in the plane of $r_A$ (the average ionic radius, related to one-electron bandwidth $W$) and $\sigma^2$ (the ionic radius variance, measuring the quenched disorder), or ``bandwidth-disorder phase diagrams'', have been established for perovskite manganites, with three-dimensional (3$D$) Mn-O network. Here we establish the intrinsic bandwidth-disorder phase diagram of half-doped layered manganites with the two-dimensional (2$D$) Mn-O network, examining in detail the ``mother state'' of the colossal magnetoresistance (CMR) phenomenon in crystals without ferromagnetic instability. The consequences of the reduced dimensionality, from 3$D$ to 2$D$, on the order-disorder phenomena in the charge-orbital sectors are also highlighted.
\end{abstract}


\maketitle

\begin{figure}
\includegraphics[width=0.46\textwidth]{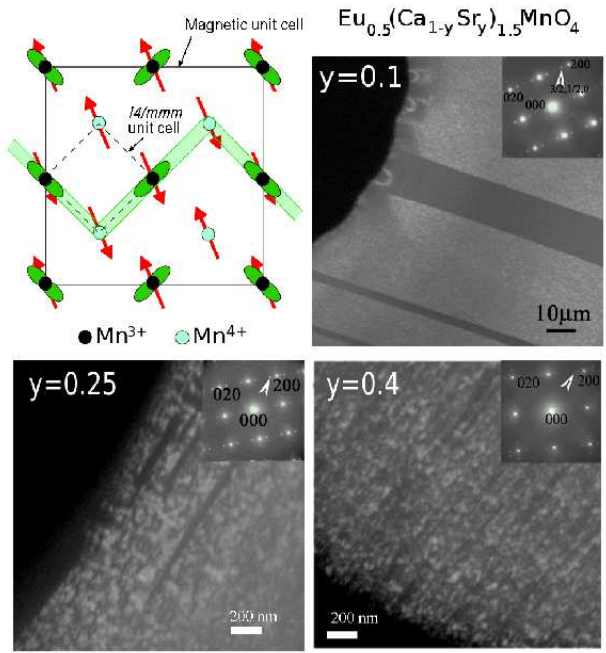}
\caption{(color online) Dark field images of Eu$_{0.5}$(Ca$_{1-y}$\-Sr$_y$)$_{1.5}$MnO$_4$, illustrating the long-range CO-OO order for $y$=0.1 and $y$=0.25 and the short-range CO-OO state for $y$=0.4; $T$ = 80K. The corresponding electron diffraction patterns (EDPs, see main text), indexed based on a tetragonal cell with $a$ $\sim$ 3.8 {\AA} and $c$ $\sim$ 12.4 {\AA} for simplicity, are shown on the right-top corners of the respective panels. A schematic view of the CE-type structure in the basal plane of the tetragonal structure is also depicted. The orbital order involves staggered $3x^2-r^2/3y^2-r^2$ orbitals of the $e_g$-like electrons of Mn$^{3+}$, represented as green (dark gray) lobes in the figure. The spins, represented with red (dark gray) arrows, order ferromagnetically along zig-zag chains, a fragment of which is highlighted in light green (light gray) in the figure.}
\label{fig-ED}
\end{figure}

Half-doped perovskite manganites with small bandwidth $W$ and small amount of disorder like Pr$_{0.5}$Ca$_{0.5}$MnO$_3$ (Pr$^{3+}$ and Ca$^{2+}$ being small and similar in size) exhibit a long-range charge and orbital order\cite{orbital, tomioka-diag} (CO-OO). This CO-OO, which is associated with the spin ordering (so-called CE-type structure\cite{jirak}), is schematically illustrated in the top-left panel of Fig.~\ref{fig-ED}. As for the spin sector, the structure is essentially composed of ferromagnetic zig-zag chains antiferromagnetically coupled to one-another. A fragment of such a zig-zag chain is highlighted in the figure. If the disorder becomes larger due to the ion size mismatch of $R^{3+}$ and $A^{2+}$, as in Gd$_{0.5}$Sr$_{0.5}$MnO$_3$, or Eu$_{0.5}$Ba$_{0.5}$MnO$_3$, only the short-range CO-OO order is observed\cite{tomioka-diag,akahoshi}, producing a ``CE-glass'' state\cite{dagotto,akahoshi,roland-ebmo}. Interestingly, the colossal magnetoresistance effect was found to arise from within this coarse-grained homogeneous CE-glass state\cite{roland-ebmo, takeshita}. In the layered systems, the MnO$_2$ planes ($ab$-planes) are isolated by two blocking ($R$/$A$)O layers, so that the CO-OO correlation is limited by the two-dimensional (2$D$) character of the Mn network. Yet, La$_{0.5}$Sr$_{1.5}$MnO$_4$ is a well-known half-doped single-layered manganite with concomitant charge and orbital ordering\cite{lsmo-orb} near 220K. The spin sector orders antiferromagnetically (AFM) at $T_N$=110K\cite{lsmo-orb}. Akin to the perovskite case, crystals with smaller bandwidth such as Pr$_{0.5}$Ca$_{1.5}$MnO$_4$ (PCMO) show CO-OO transitions above room temperature\cite{raveau}. However, no other half-doped RSMO system seems to exhibit a long-range CO-OO. A CE-glass state is observed in crystals with larger quenched disorder, such as Eu$_{0.5}$Sr$_{1.5}$MnO$_4$\cite{roland-esmo} (Eu$^{3+}$ is smaller than La$^{3+}$, which is already smaller than Sr$^{2+}$). In the present article, using high-quality single crystals of $R_{0.5}A_{1.5}$MnO$_4$ manganites, we investigate the CE-glass state and its location in the plane of quenched disorder vs. bandwidth. The quenched disorder associated with the solid solution of the $A$-site cations\cite{akahoshi} is quantified using the ionic radius variance $\sigma^2=\sum_i x_i r_i^2 - r_A^2$, according to the scheme devised by Attfield\cite{attfield}. $x_i$ and $r_i$ are the fractional occupancies ($\sum_i x_i$=1) and electronic radii of the different $i$ cations on the $A$-site, respectively, and $r_A=\sum_i x_ir_i$ represents the average $A$-site ionic radius, related to the bandwidth.

High quality single crystals of the $A$-site disordered $R_{0.5}$Ca$_{1.5}$MnO$_4$ (RCMO), $R_{0.5}$Sr$_{1.5}$MnO$_4$ (RSMO), and  $R_{0.5}$(Ca$_{1-y}$Sr$_y$)$_{1.5}$MnO$_4$ (RCSMO) manganites were grown by the floating zone method ($R$ = La, La$_{1-y}$Pr$_{y}$ , Pr, Nd, La$_{0.5}$Eu$_{0.5}$  ($\sim$ Nd), Sm, or Eu, while $A$=Ca, Ca$_{1-y}$Sr$_y$). The phase-purity of the crystals was checked by x-ray diffraction and the cation concentrations of some of the crystals were confirmed by inductively coupled plasma (ICP) spectroscopy. The ac-susceptibility $\chi$($\omega=2\pi f$) data was recorded as a function of the temperature $T$ and frequency $f$ on a MPMSXL SQUID magnetometer equipped with the ultra low-field option (low frequencies) and a PPMS6000 (higher frequencies) from Quantum Design, after carefully zeroing or compensating the background magnetic fields of the systems. The resistivity $\rho$ of the crystals was measured using a standard four-probe method on a PPMS6000, feeding the electrical current in the $ab$-plane. The single-crystal x-ray data was recorded at 370K on a Rigaku SPD curved imaging plate system at the beam line BL-1A of the Photon Factory, KEK, Japan. Thin specimens were prepared for observation with transmission electron microscopes (TEMs) by Ar$^+$ ion milling at low temperatures, to perform the electron diffraction (ED) measurements, and collect the selected-area electron diffraction patterns (EDPs) and dark-field (DF) images. The structural modulation wave vector $q$= $a^*$$[\delta$ $\delta$ $0]$ ($a$ is the lattice constant, $aa^*$=1) was determined at different temperatures.

Eu$_{0.5}$Ca$_{1.5}$MnO$_4$ (ECMO) is very similar to PCMO, albeit a larger variance ($\sigma^2$ $\sim$ 7$\times$10$^{-4}$ {\AA}$^2$ instead of $\sim$ 2$\times$10$^{-7}$ {\AA}$^2$ for PCMO). The CO-OO remains long-ranged in all the RCMO crystals, even when a small amount of Ca is substituted with Sr. For example, in the insets of Fig.~\ref{fig-ED}, we show the [001] zone-axis electron diffraction (ED) patterns of Eu$_{0.5}$(Ca$_{1-y}$Sr$_y$)$_{1.5}$MnO$_4$ (ECSMO)  obtained at 80 K. In addition to the fundamental spots (associated with the K$_2$NiF$_4$ structure), the EDPs include superlattice (SL) spots, associated with the CO-OO. The sharpness and the modulation wave vector however are dependent on the Sr concentration (see below). The different panels of  Fig.~\ref{fig-ED} illustrate the changes in the microstructure related to the CO-OO with increasing Sr content. These dark-field (DF) images were recorded at 80K, using the SL reflection marked by the arrow in the electron diffraction patterns (EDPs). The bright regions in Fig.~\ref{fig-ED} correspond to regions where the CO-OO occurs. For $y$=0.1 large CO-OO domains are observed.  On increasing Sr content, the size of the CO-OO domains decreases ($y$=0.25), until the CO-OO becomes short-ranged ($y$=0.4).

\begin{figure}
\includegraphics[width=0.46\textwidth]{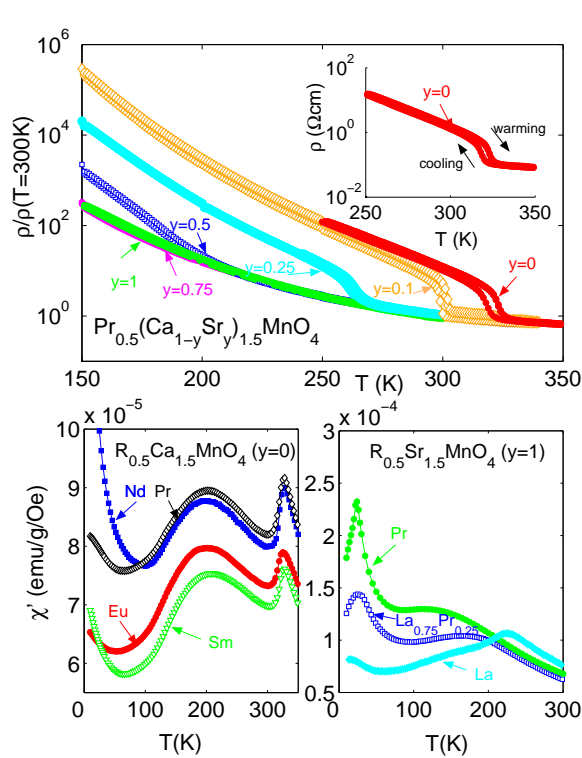}
\caption{(color online) Top panel: Temperature $T$ dependence of the normalized in-plane resistivity $\rho/\rho_{(T=300 K)}$ for Pr$_{0.5}$(Ca$_{1-y}$Sr$_y$)$_{1.5}$MnO$_4$. The inset shows the resistivity $\rho$ of the crystal with $y$=0 (PCMO) in absolute units. Lower panel: Temperature dependence of the in-plane component of the ac-susceptibility $\chi'$ for some of the (left) $R_{0.5}$Ca$_{1.5}$MnO$_4$ and (right) $R_{0.5}$Sr$_{1.5}$MnO$_4$ crystals. The low-temperature upturn of $\chi'$($T$) in the RCMO crystals is attributed to the 4$f$ moments of the $R$ cations.}
\label{fig-XRT}
\end{figure}

The order-disorder in the charge-orbital sector also affects macroscopic properties such as the magnetization or ac-susceptibility, as well as the electrical resistivity. For example, the disappearance of the long-range CO-OO state is observed in the $T$-dependence of the electrical resistivity, as shown in the upper panel of Fig.~\ref{fig-XRT} for Pr$_{0.5}$(Ca$_{1-y}$Sr$_y$)$_{1.5}$MnO$_4$ (PCSMO). The $\rho(T)$ curves show a clear (and hysteretic in temperature) inflection near the CO-OO transition temperature $T_{\rm CO-OO}$ up to $y$=0.5,  for which no CO-OO phase transition occurs, as confirmed by the ED data. $T_{\rm CO-OO}$ is also clearly observed in the $\chi(T)$ curves as the sharp peak arising from the quenching of the FM spin fluctuation. Figure~\ref{fig-XRT} shows the temperature dependence of the in-phase component of the ac-susceptibility $\chi'$ for some of the single crystals. $T_{\rm N}$ is however difficult to identify, as seen for example in the $\chi(T)$ curves of the well known LSMO\cite{lsmo}. As seen in the left lower panel of Fig.~\ref{fig-XRT}, in the RCMO crystals with small disorder ($\sigma^2$ $<$ 1$\times$10$^3$ {\AA}$^2$) and relatively small average ionic radius ($r_A$ $\sim$ 1.16-1.18 {\AA}), a sharp peak marking  $T_{\rm CO-OO}$ is observed above 320K. At lower temperatures, near 200K, a broader peak is observed. This broader peak does not correspond to $T_{\rm N}$ for long-range spin order, which is $\sim$ 120-130K in these crystals\cite{Jing}. An inflection (more clearly seen in the $T$-derivative of $\chi'(T)$) can been seen in the vicinity of these temperatures, which was found to coincide with the $T_{\rm N}$ determined by diffraction techniques\cite{Jing}. 
The broad maximum near 200K may thus indicate the development of in-plane AFM correlation, rather than the long-ranged phase AFM transition. We refer in the following to this broad peak as $T_{\rm S}^*(ab)$. In the $R$SMO crystals, with larger $r_A$ ($\sim$ 1.28 {\AA}) and bandwidth, only the susceptibility of LSMO (the right lower panel of Fig.~\ref{fig-XRT}) shows the $T_{\rm CO-OO}$ peak, as well as a bump near 150K which may reflect the above mentioned in-plane spin correlation. As the variance (quenched disorder) increases with substitution of the La ions with Pr, only a broad peak is observed at high temperatures, together with a broad frequency-dependent cusp at low temperatures\cite{newnote2}. We now compare these observations with the electron diffraction data. In the EDPs collected as a function of temperature, the superlattice spots associated with the CO-OO are observed for all the RSMO crystals (one such spot is marked with an arrow in the EDP of ECMO shown in the corner of the top-left panel of Fig.~\ref{fig-ED}). However, these SL spots are sharp only for LSMO, and diffusive, more or less, for the crystals with larger $R$, confirming the short-ranged nature of the CO-OO correlation\cite{uchida} as suggested by the absence of a sharp peak in $\chi(T)$. Thus in the half-doped case, the orbital sector, as the master, controls the spin sector, as the slave, determining the spatial extent of its correlation as well. 

\begin{figure}
\includegraphics[width=0.46\textwidth]{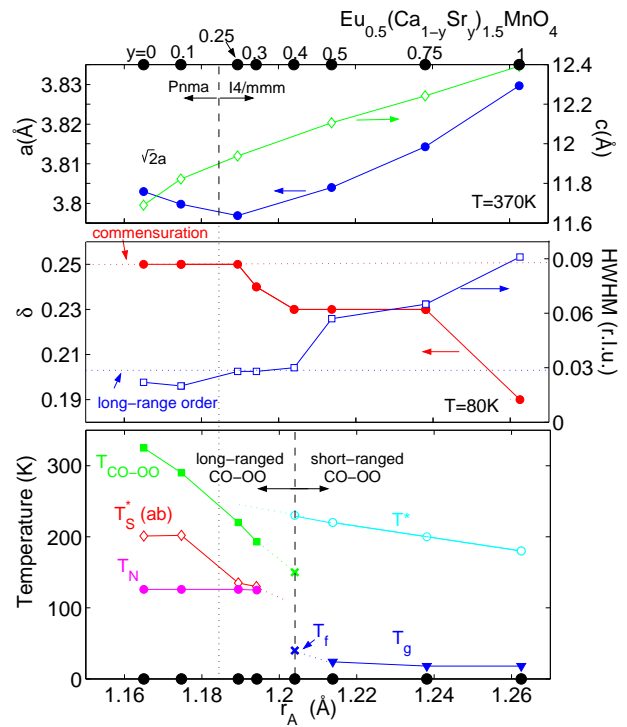}
\caption{(color online) Average ionic radius $r_{\rm A}$dependence of selected physical properties of Eu$_{0.5}$(Ca$_{1-y}$\-Sr$_y$)$_{1.5}$MnO$_4$. Top: $a$- and $c$-axis lattice parameters; $a$ $\sim$ $b$ for the crystals with $Pnma$ structure. Middle: Modulation wave vector $\delta$ and half-width at half-maximum HWHM of the superlattice reflection intensity profile obtained from electron diffraction (ED) at $T=80$K. Bottom: the electronic phase diagram of Eu$_{0.5}$(Ca$_{1-y}$Sr$_y$)$_{1.5}$MnO$_4$ (see main text for the definitions of the different labels). The crosses mark features in $\rho(T)$ or $\chi(T)$ curves, which do not necessarily correspond to phase transitions.}
\label{fig-all}
\end{figure}
The distinction between long-range and short range CO-OO is investigated in more detail, as a function of the bandwidth ($r_A$, or the Sr concentration $y$) in Fig.~\ref{fig-all}. The top panel of Fig.~\ref{fig-all} shows the variation of the lattice parameters of ECSMO, estimated at high temperatures ($>$ $T_{\rm CO-OO}$) from the single-crystal x-ray diffraction. The $a$- and $c$-axis parameters decrease significantly with decreasing $y$, down to $y$=0.25. These crystals have a tetragonal $I4/mmm$ structure similar to those of the RSMO crystals. For $y$ $<$ 0.25, the structure is orthorhombically distorted\cite{newnote2}. However, this structural transition does not coincide with the appearance of the long-range CO-OO order. The $\chi(T)$ and $\rho(T)$ curves suggest that the CO-OO order becomes short-ranged near $y$=0.4. This is confirmed by the ED data, as illustrated in the middle panel of Fig.~\ref{fig-all}.
\begin{figure}
\includegraphics[width=0.46\textwidth]{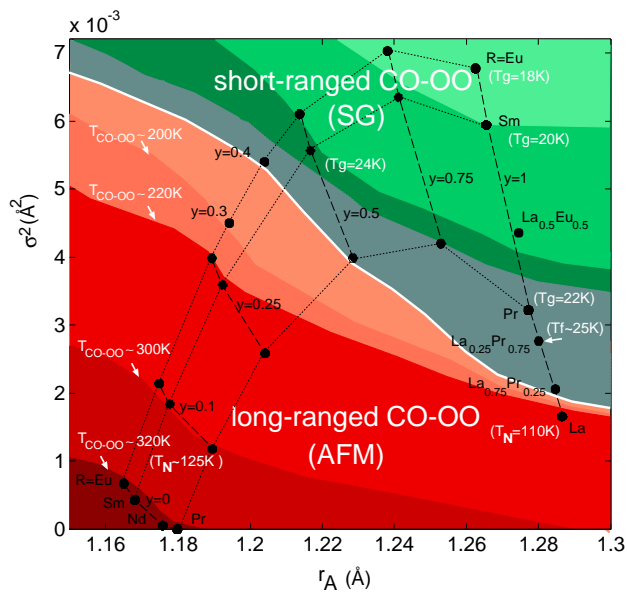}
\caption{(color online) Electronic phase diagram of $R_{0.5}$(Ca$_{1-y}$Sr$_y$)$_{1.5}$MnO$_4$ in the plane of the average ionic radius $r_A$ and the variance $\sigma^2$. Dashed lines connect the crystals with the same Sr content $y$, while dotted line connect crystals with the same $R$ cation. Data of the CO-OO (the CO-OO transition temperature $T_{\rm CO-OO}$) and magnetic (the spin-glass (SG) phase transition temperature $T_g$, the freezing temperature $T_f$, and the antiferromagnetic (AFM) transition temperature $T_{\rm N}$, in parenthesis) are included.}
\label{fig-diag}
\end{figure}
The half-width at half-maximum (HWHM) of the CO-OO superlattice spots in the EDPs is proportional to the inverse of the CO-OO correlation length $\xi_{\rm CO-OO}$. In the case of $y$=0.4, the HWHM is relatively small, however dark-field imaging reveal the short-ranged nature of the CO-OO order (c.f. Fig.~\ref{fig-ED}). As $y$ increases above 0.4, the HWHM gradually increases, i.e. $\xi_{\rm CO-OO}$ gradually decreases, down to the nanometer-scale\cite{roland-esmo}. The modulation wave vector of the SL spots also varies with $r_{\rm A}$ (or $y$):  it is commensurate ($\delta$=1/4) up to y=0.25, and becomes incommensurate for $y$=0.3, although the CO-OO order is still long-ranged. It remains incommensurate for $y$ $\geq$ 0.4 (short-range CO-OO order). The variation of the magnetic, and electrical properties of ECSMO are summarized in the bottom panel of Fig.~\ref{fig-all}. $T^*$ marks the appearance of diffuse superlattice spots in the EDPs (i.e. CO-OO correlation)\cite{uchida}, while $T_g$ is the SG phase transition temperature, obtained from the dynamical scaling of the $T_f(f$) freezing data of $\chi(T,f)$\cite{roland-ebmo,roland-esmo}. In the case of the crystal with $y$=0.4, no true SG phase transition is found, albeit glassiness below $T_f$ $\sim$ 40K. While $T_{\rm CO-OO}$ largely varies with $r_A$, $T_{\rm N}$ is relatively unchanged for all the crystals with long-range CO-OO order.

The schematic phase diagram presented in Fig.~\ref{fig-all} can be also plotted as a function of $\sigma^2$. The resulting diagram is very similar, as both $r_A$ and $\sigma^2$ vary significantly with $y$. It hence makes sense to draw a global phase diagram in the planes of $r_A$ and $\sigma^2$ to take into account the effects of the variation of both bandwidth and quenched disorder. Such a ``bandwidth-disorder'' phase diagram is drawn in Fig.~\ref{fig-diag}, using the ac-susceptibility, resistivity, and electron diffraction data, which was found to complement each other in the above. This phase diagram is reminiscent of the diagram obtained for the 3$D$ perovskite case\cite{tomioka-diag} in the small bandwidth area (for larger $W$, FM is observed in the perovskite case). In both cases, the long-range CO-OO order is replaced by a short-range ``CE-glass'' state  (SG state) in the presence of large quenched disorder \cite{roland-ebmo,roland-esmo}. However, the first-order like transition between the CO-OO and CE-glass phases observed in the perovskite case\cite{akahoshi} does not occur in the layered systems. As indicated by the ED results, the CO-OO correlation length continuously decreases as the quenched disorder increases. Since there is a clear covariation between the CO-OO correlation length and size of the "superspins" involved in the SG state, the latter of which was determined by the dynamical scaling of the $\chi(T,f)$ data\cite{roland-esmo,uchida}, these groups of coherent spins may be viewed as broken pieces of the CO-OO FM zig-zag chains of the CE-type structure.

To summarize, we have established for the first time the intrinsic bandwidth-disorder phase diagram of the half-doped layered manganites using high-quality single-crystals. As in the perovskite case, the CE-glass state occupies a large area of the diagram. Many specimens were found to exhibit a long-range CO-OO, with a $T_{\rm CO-OO}$ tunable around room-temperature and above by the bandwidth and/or disorder. The macroscopic phase separation, or ferromagnetic phases, sometimes reported in studies on polycrystals is not observed. Remarkably, the present diagram is very similar to that of the narrow-bandwidth perovskites, in spite of the dimensionality difference. However, in the present 2$D$ layered case, the gradual decrease of the CO-OO correlation length as a function of bandwidth or disorder occurs, instead of the first-order-like collapse observed in the 3$D$ case.


\begin{thebibliography}{1}
\bibitem{newnote}
The broad peak at high-$T$ vanishes near $R$=Eu, which shows a sharp cusp at low $T$. The crystals with long-range CO-OO have no $f$-dependent $\chi'(T)$, and $\chi''(T)$ is negligible. On the other hand, the crystals with short-range CO-OO show some glassiness or well-defined SG phase transitions related to the sharp cusps in $\chi'(T)$.
\end{thebibliography}
\end{document}